\documentclass[nohyper,12pt,letterpaper]{JHEP}

\usepackage{epsfig}

\def \eqn#1#2{\begin{equation}#2\label{#1}\end{equation}}

\title{Some Thoughts on the Quantum Theory of de Sitter Space}
\author{T.\,Banks\\
  SCIPP, University of California, Santa Cruz, CA 95064\\
{\it and}\\
NHETC, Rutgers University, Piscataway, NJ 08854\\
  E-mail: \email{banks@scipp.ucsc.edu}}

\abstract{This is a summary of two lectures I gave at the Davis
Conference on Cosmic Inflation. I explain why the quantum theory
of de Sitter (dS) space should have a finite number of states and
explore gross aspects of the hypothetical quantum theory, which
can be gleaned from semiclassical considerations.  The constraints
of a self-consistent measurement theory in such a finite system
imply that certain mathematical features of the theory are
unmeasurable, and that the theory is consequently mathematically
ambiguous.  There will be a universality class of mathematical
theories all of whose members give the same results for local
measurements, within the {\it a priori} constraints on the
precision of those measurements, but make different predictions
for unmeasurable quantities, such as the behavior of the system on
its Poincare recurrence time scale.  A toy model of dS quantum
mechanics is presented.}

\keywords{Quantum Gravity, de Sitter Space}

 \received{} \accepted{}

\preprint{SCIPP-03/04, RUNHETC-2003-12}

\begin{document}

%%%%%%%%%%%%%%%%%%%%%%%%%%%%%%%%%%%%%%%%%%%%%%%%%%%%%%%%%%%%%%%%%%%%%%%%%%%%
%          Table of contents automatic
%%%%%%%%%%%%%%%%%%%%%%%%%%%%%%%%%%%%%%%%%%%%%%%%%%%%%%%%%%%%%%%%%%%%%%%%%%%%

\section{\bf Introduction}

Observations suggest that the expansion of the universe is
accelerating.  The simplest explanation of this acceleration is a
non-zero cosmological constant.  Conventional string/M theory
cannot accommodate a positive cosmological constant.   At best it
can have metastable positive energy density minima of an
approximately defined effective potential in extreme regions of
moduli space\cite{kklt}.  The significance of these minima is
currently under study.  In these talks I concentrated on the
alternative hypotheses that there is an extension of string theory
which defines quantum theor(ies) of gravity for any positive
cosmological constant, $\Lambda$, which approach an isolated
super-Poincare invariant string/M vacuum in the limit of vanishing
$\Lambda$.  The finite $\Lambda$ theories are
conjectured\cite{tbfolly}\cite{willy} to have a finite number of
physical states.

The fact that different values of the cosmological constant define
different quantum systems, rather than different superselection
sectors of the same system, is, for negative cosmological
constant, one of the predictions of the AdS/CFT correspondence.
One can view the negative cosmological constant as a parameter
which controls the behavior of the high energy density of states
in an asymptotically AdS universe.   For positive cosmological
constant the conjecture is that there is a high energy cutoff, and
a finite number of states, controlled by the value of $\Lambda$.

To understand why one would make such a conjecture, consider the
description of dS space in terms of asymptotic past and future
boundaries ${\cal I}_{\pm}$, as advocated by Witten\cite{edindia}
and Strominger\cite{dscft}.   Naively, one might imagine a theory
of correlation functions on these boundaries, which could be
interpreted as defining a sort of S-matrix for dS space.   In the
semiclassical approximation, these would be defined by solutions
of the bulk field equations, with boundary conditions on ${\cal
I}_{\pm}$ .  In asymptotically flat or AdS universes, such a
definition makes sense because generic solutions with such
boundary conditions exist, and have at most localized
singularities\footnote{If the Cosmic Censorship hypothesis is
correct, every such singularity is shrouded behind a black hole
horizon.}.  The phase space of the system in these cases is indeed
parametrized by arbitrary perturbations on the boundary.

This is not correct in asymptotically dS space.   Heuristically,
if we insist that each boundary probe insert a minimal finite
energy density (in global coordinates), no matter how small, into
the system, then the solution with some finite number of probes
will have an energy density of order the Planck scale by the time
the minimal radius of dS space is achieved.   The solution will
have a singular Big Crunch or Big Bang and will fail to be
asymptotically dS in either the past or the future\footnote{The
question of whether such singular, unidirectionally asymptotically
dS spaces have a finite number of states is more complicated. To
answer it one must have a theory of the singularity.  If one makes
the hypothesis that solutions in which the singularity occurs at
different values of the radius of the dS sphere should not be
included in the same system, then these spaces have only finite
numbers of states as well.}. A more mathematical statement of this
is that the phase space of Einstein gravity plus sensible matter
with asymptotically dS boundary conditions in both past and future
is compact.  This is not yet a theorem, but is believed by many
relativists. Preliminary results in this direction were obtained
in unpublished work of Horowitz and Itzhaki\cite{HI}.  It is well
known that classical systems with compact phase space have a
finite number of states when quantized.

A different, but related argument for a finite number of states
comes from a study of classical and semiclassical physics within
the causal diamond of a timelike observer in dS space.  This
region has a timelike Killing vector, and the corresponding
classical Hamiltonian \cite{gt} has finite energy excitations
corresponding to Kerr-dS black holes of various radii and angular
momenta.  There is a maximum energy black hole: the Nariai
solution.   Semiclassical arguments\cite{gh} indicate that the
density matrix describing dS space is thermal, with a fixed
temperature and finite entropy.  When combined with an upper bound
on the energy spectrum, a finite entropy thermal density matrix
implies a finite number of states.   This was the original
argument in \cite{tbfolly}.  The two arguments are related because
the result of a too violent perturbation of the boundary of dS
space is undoubtedly the creation of large black holes.  When the
black hole size exceeds the Nariai limit, we instead get a
spacelike singularity covering an entire Cauchy surface of the
space-time.

\section{\bf Groups, and boundaries}

Isometries are diffeomorphisms and diffeomorphisms are gauge
transformations.  Thus, in general we do not expect an action of
isometries on the physical states of a quantum system which obeys
the principle of general covariance.  Instead, global symmetry
groups are constructed from equivalence classes of diffeomorphisms
with certain action on the timelike or null boundary of a
space-time.  Global dS space does not have such boundaries, but
the causal patch of a timelike observer does.  In \cite{bfmcosmo}
it was suggested that the entire quantum mechanics of dS space
could be understood from the point of view of any given observer,
with the different descriptions related by a Cosmological
Complementarity principle.  This is a generalization of the idea
of Black Hole Complementarity introduced by Susskind, Uglum and
Thorlacius and by 't Hooft.  E. Verlinde has suggested the name
Observer Complementarity for the general principle which states
that observers who are classically causally disconnected, use the
same Hilbert space in the quantum theory (in field theory they
would use independent tensor factors of the Hilbert space), but
measure mutually non-commuting observables.

Gomberoff and Teitelboim\cite{gt} have shown how to construct
classical symmetry generators for general Kerr-dS black hole
spacetimes.  In this formalism one can treat empty dS space as the
zero mass limit of a black hole.   The only generators which make
sense are those which preserve the causal patch and its
cosmological horizon, which is the surface on which boundary
conditions are imposed and the generators are defined.  These form
an $R \times SO(3)$ subgroup of the $SO(4,1)$ isometry group of
$dS_4$.  The coset of this subgroup in the dS group is a set of
gauge transformations, which map the physical Hilbert space into
copies of it viewed from the point of view of different timelike
observers.

This restriction to a subgroup is satisfying, because it is
compatible with a quantum theory with a finite number of physical
states.  However, it seems to raise a puzzle about how the
Poincare group will emerge in the small $\Lambda$ limit, since we
usually realize it as a contraction of the full dS group. In
order to understand what is going on, we should compare the
boundaries of the two space-times, where the symmetry generators
are defined, rather than the bulk, where the isometries act as
diffeomorphisms {\it i.e.} gauge transformations.

The near horizon geometry of dS space is

\eqn{nearhor}{ds^2 = R^2 (du dv + d\Omega^2)}

\noindent where $v \rightarrow 0$ defines the future cosmological
horizon. The static Hamiltonian is the boost generator on $u$ and
$v$.

Future null infinity in asymptotically flat spacetime is the
$v\rightarrow 0$ limit of

\eqn{nullinf}{ds^2 = {(du dv + d\Omega^2 )\over v^2}}.

To remove the infinity one performs an infinite conformal
transformation, and obtains a manifold with only conformal
structure and coordinates $(u, \Omega )$.  The asymptotic
symmetry group is the semi-direct product of the conformal group
of the sphere (isomorphic to the Lorentz group) with the infinite
abelian group whose generators are $f(\Omega ) \partial_u$, with
arbitrary $f$ . This is the Bondi-Metzner-Sachs group.  If we
restrict attention to classical vacuum spacetimes with no
classical gravitational radiation (as would be appropriate in
dimensions where an S-matrix exists), this can be reduced to the
Poincare subgroup where $f$ is restricted to be the constant
function or the $j=1$ spherical harmonic.

Apart from the rotations of the sphere, the static dS subgroup and
the Poincare group have no generators in common.  Lorentz
invariance is a conformal symmetry which appears only in the limit
of infinite dS radius.

Thus, in constructing a quantum theory of dS space, one is
interested in finding two different Hamiltonians.  One is an exact
symmetry which describes time translation for a static observer at
the center.  The second is the Poincare Hamiltonian, which is
relevant for describing scattering processes in the limiting
asymptotically flat spacetime.   {\it A priori} there is no
connection between these two generators, but we might expect some
similarities for a subclass of localizable low energy states.

\section{\bf States, localization, and temperature}

The local rule for maximizing entropy in a gravitational system
is to form a large black hole.  In dS space this procedure reaches
a limit at the Nariai black hole.  Every black hole in dS space
has both a cosmological and a black hole horizon, but even the sum
of the entropies of the two horizons of the Narai black hole is
only $2/3$ of the entropy of the full dS space.  Indeed, the total
entropy of a black hole monotonically decreases in dS space, while
the entropy associated with the black hole horizon increases.

This is an indication that the black hole states are actually
relatively low entropy states of the full dS spacetime.  They are
made by borrowing degrees from the horizon of empty dS space and
freezing them into special configurations.  We will see a rather
explicit model of this below.

Another class of states in dS quantum mechanics may be termed
field theoretic states.  These are the states which are well
approximated by the treatment of quantum field theory in curved
spacetime.   Their back reaction on the geometry is supposed to be
negligible.   At the very least, we must require that they do not
collapse into black holes.

The entropy of field theory states in a given volume (with no
constraint on the energy) is dominated by the ultraviolet, where
the field theory is described by a fixed point.   The entropy is
of order $M^3 R^3$, where $M$ is a UV cutoff.   The energy of
these states is of order $M^4 R^3$.  In order that the
Schwarzchild radius of the system be less than $R$, we must have
$M^4 R^2 < 1$, in Planck units.  Thus, the entropy in field
theoretic states is less than $R^{3\over 2}$.   In the static
patch of dS space, $R$ is at most the dS radius.  Thus, field
theoretic states are even less entropic than black holes.

One concludes that most of the states are viewed by the static
patch observer as being associated with the horizon of empty dS
space.   These have classical energy $0$.   From the point of view
of classical GR, the association of the bulk of the entropy with
the horizon is quite reasonable.  The static observer never sees
anything fall through the horizon, but does see any object to
which she is not bound, get squeezed into an infinitesimal region
near the horizon.  The global observer sees these objects as
living in a large set of disjoint horizon volumes, and finds a
natural explanation for most of the entropy being invisible to any
given static observer .

Our identification of field theoretic states allows us to
understand how the global point of view of quantum field theory in
curved spacetime might be approximately correct.  Above we have
stressed Observer Complementarity:  all of the physics can be
described from the point of view of a given static observer. In
the analogous, but different situation of black holes in
asymptotically infinite spaces, none of the degrees of freedom
associated to the black hole horizon by the Schwarzschild
observer, commute with the complete set of asymptotic
observations made by this observer.  If they did, then the
principle of Black Hole Complementarity would not resolve the
Information Paradox. All of the information available to the
Schwarzschild observer can be read by him in terms of scattering
measurements which are describable by local field theory  (that
is, the measurement itself, not necessarily the scattering
amplitude being measured, can be so described).

The static observer in dS space is, in some respects, analogous to
the Schwarzschild observer, but we have identified an important
difference.  Only a fraction of order $R^{-{1\over 2}}$ of the
total entropy available in the static observer's Hilbert space
refers to local field theoretic measurements.  This is consistent
with the possibility of making of order $R^{1/2}$ commuting copies
of the field theoretic degrees of freedom in a given horizon
volume.  Quantum field theory in global dS space predicts that in
the far future one can find an infinite number of commuting copies
of the degrees of freedom in a given horizon volume.  Here we see
that this becomes a better approximation at large $R$, although
the number of copies is never strictly infinite.

If on the other hand, we make black holes whose size scales with
the horizon, then there is no similar multiplication.   The full
quantum theory of dS space must be able to accommodate both sorts
of state.  However, it is amusing to note that the field theoretic
states {\it in a maximal collection of disjoint horizon volumes}
seem to be more typical states of the system than those with large
black holes.  They can saturate the full dS entropy.

The states which a static observer associates with the horizon of
empty dS space dominate the thermal density matrix at the dS
temperature.  This is surely true if they have strictly zero
energy.  A much more likely picture is that they are distributed
between $E = 0$ and a cutoff $E=\Delta$, with a density $e^{-
S_{dS}}$.  In this case they will still dominate the thermal
entropy.  Higher energy states are much less entropic.   As a
consequence, the thermal entropy of dS space will be close (in the
limit of large $R$) to the logarithm of the number of horizon
states, which, in the same limit, is approximately the total
number of states of the system.

This picture of the spectrum of the dS Hamiltonian, suggests a
self consistent explanation for the origin of the dS temperature.
Namely, if we postulate this dense set of low energy states, and
assume that their Hamiltonian is a random matrix, so that dynamics
in this subspace is chaotic (as chaotic as a finite quantum system
can ever be), then perhaps we need only postulate a weak coupling
between these states and any other states of the system in order
to explain why the localized systems experience thermal
fluctuations.  The dS temperature would then be determined by the
cutoff $\Delta$ on the low energy spectrum.  My student, Lorenzo
Mannelli, is trying to prove this.

\section{\bf Measurement theory in dS space}

Theoretical physics was invented to describe the result of outside
measurement on an isolated system.  With the advent of quantum
mechanics, we have had to pay a little more attention to what we
really mean by a measuring process.  Initial discussions of this
had to assume the existence of a separate classical world of
measuring equipment.   More modern discussions view this as an
approximate description of a self consistent process of
measurement of one quantum system by another.  The discussion in
this section of measurement theory in dS space is based on the
paper\cite{nightmare}.

There has been much discussion in the recent measurement theory
literature of ``environmental decoherence": The effects of random
interactions between the measurement apparatus and a large
unmeasured ``environment".  While not denying the existence of
such effects for all realistic measurements, I would like to
believe that they are not logically necessary to the existence of
a sensible theory of measurement.  If we are to take the step of
extending the formalism of quantum theory to describe the entire
universe, we must give up the crutch of unmeasured environments.

I believe that a reasonable measurement theory exists, without
postulating environmental decoherence.  All of measurement theory
rests on Von Neumann's observation that ordinary unitary evolution
can take an uncorrelated state of a system plus a measuring
apparatus into an entangled state in which each eigenstate of a
complete set of commuting observables of the system, is correlated
with a different ``pointer" state of the apparatus.

\eqn{measeq}{\sum a_n |n> |A> \rightarrow \sum a_n |n> |A_n>}

In the theory of environmental decoherence, it is assumed
properties of the pointer states' interaction with the random
environment that enable one to claim that further measurements on
the system will not be sensitive to the relative phases of the
$a_n$.  An alternative explanation of decoherence is illustrated
by a simple model.  Suppose we are trying to measure a single
spin, and we model our measuring apparatus by a cutoff quantum
field theory with two degenerate minima, $\phi_{\pm}$ in a volume
$V$ which is large in cutoff units.   Postulate a nonlocal
coupling of the spin to the field theory which correlates the
state $\sigma_3 = 1$ with $\phi_+$ and $\sigma_3 = -1 $ with
$\phi_-$.   This is a cartoon of the amplification that is
necessary to get a microscopic phenomenon to register on a
macroscopic apparatus.

What do we mean by this correlation?  $\phi_{\pm}$ are not really
single states but labels for whole ensembles of states in which
the field takes values very close to $\phi_{\pm}$ in most of the
volume $V$.  Now consider any operator $O_{loc}$ which is
localized in a volume much less than $V$.   The matrix elements of
$O_{loc}$ between any pair of states from the two different
ensembles, is of order $e^{-V}$.  I now claim that our correlated
state is one in which we can say that a measurement has been made.
Further local perturbations of the apparatus will not change the
fact that the states where the spin is positive and negative can
communicate only by amounts of order $e^{-V}$.   Expectation
values of system operators in the correlated state and all states
it evolves into under local perturbations over times short
compared to $e^V$ will follow the rules of classical probability.

Over times of order $e^V$, tunnelling between the two would be
superselection sectors will occur, and the measurement will lose
its coherence.  But in ordinary quantum mechanics we can imagine
taking $V$ as large as we like.   Thus we can approximate
Copenhagen measurements of quantum systems as well as we like, and
thus give operational meaning to the mathematically precise
formulae of the quantum theory.

In theories of quantum gravity this argument must be rethought.
The large, almost classical, measuring devices will gravitate and
have potentially large effects on the system they are supposed to
be measuring.  The only way to avoid this, is to place the
measuring devices further and further away from the system, as we
try to make them larger in order to make the measurements more
precise and more robust against quantum fluctuations in the
apparatus.  This is why the only mathematically precise
observables in good theories of quantum gravity are S-matrix
elements (and their analogs in other infinite geometries).

We can see that when we come to dS space we are in a bind. If we
are trying to measure the results of an experiment, which is
traveling along a particular timelike geodesic, the best we can
do is to measure its influence on a freely falling detector that
is practically at the cosmological horizon of the experimental
system.  This is the closest analog of a scattering matrix that
can be achieved in dS space.  The detector can be made very large
without significant effect on the experiment.

The key question now is how large it can be.   If we require that
the detector's workings can be understood with ``current
technology", then according to the above discussion, the detector
must be built from what we have called field theoretic states in
the static patch.  In that case, an extremely conservative lower
bound on the tunneling amplitudes between pointer states of the
detector, is of order $e^{- b R^{3/2}}$, with $b$ a constant
(much) less than one and $R$ the radius of dS space.   This can
only be achieved with detectors whose size is a finite fraction of
the cosmological horizon.

There is a hypothetical possibility for the construction of more
robust detectors.  For field theoretic detectors, tunneling
amplitudes, are of order $N^{-p}$, where $p$ can be a number of
order $1$, and $N$ is the number of states of the detector. Black
holes whose size scales like the cosmological horizon have a much
larger number of states than any field theoretic system.  In
principle they could provide the mechanism for more robust
detectors.  However, in order for that to work, one must be able
to construct pointer states for the black hole.  In field
theoretic models the robustness of pointer states depends on the
concept of superselection sector, which is itself a consequence of
locality.  Such considerations do not apply to the states on a
black hole horizon.  Indeed, the existence of an elaborate set of
pointer states of a black hole, which would enable us to make
precise and robust measurements of a multitude of observables
external to the black hole, would seem to contradict the no hair
theorem and the thermal nature of black hole physics. Nonetheless,
since we cannot rule out the possibility rigorously, the use of
black holes as detectors must be considered.  The tunneling
amplitudes between pointer states of such monstrous detectors
would be bounded from below by something of order $e^{- c R^2}$.
Again we would expect $c$ to be much less than one.

These considerations imply that there is a fundamental limit, both
to the precision of any measurement in dS space and to the amount
of time for which any actual physical object in dS space can play
the role of an idealized Copenhagen measuring device.   {\it This
time scale is always much less than the Poincare recurrence time,}
even if we accept the bizarre possibility of detectors constructed
from the microstates of black holes.

Historically, mathematical formulae for observable quantities in
theoretical physics were presumed (to the extent they were
presumed exactly correct) to be precise results to which actual
measurements could approximate with any required degree of
precision.  Once we accept the rules of quantum mechanics, and the
hypothesis that the entire universe has a finite number of
physical states this can no longer be correct.  Considerations of
gravitational interactions and the geometry of dS space give us a
more refined estimate of the fundamental limits on the precision
of measurements in such a situation.  It seems absolutely clear
that there will then be {\it many} Hamiltonian descriptions of the
physics of dS space, that will fit all conceivable experiments
within the fundamental limits on their precision.  It also seems
clear where the modifications that do not affect ordinary
measurements will come from.   Most of the states on the horizon
do not affect measurements in the interior, apart from providing
the thermal bath at the dS temperature, and perhaps renormalizing
the effective local field theory Lagrangian describing field
theory states in the interior\cite{sushor}.  We have already
suggested the idea that the horizon states could be described by a
random Hamiltonian with an appropriate spectral cutoff related to
the dS temperature.

It seems likely to me that the proper mathematical description of
this situation will utilize the concept of universality classes
from the theory of phase transitions.   dS space is a finite
system.  The vanishing cosmological constant limit is a critical
limit in which the number of states goes to infinity.  There will
be a universality class of Hamiltonians which describe dS space in
this limit, and give the same answers for all observables with
the fundamental limits on precision that we have outlined.  Our
considerations suggest that the predictions of these different
mathematical theories will be the same, over reasonable periods of
time, to all orders in powers of the cosmological constant.  The
imprecisions we have identified vanish like the exponential of a
power of the cosmological constant.    This means that for all
practical purposes, the mathematical formulation of dS space will
be predictive.

However, when it comes to questions of what happens to the system
over a Poincare recurrence time \cite{susetal} the different
Hamiltonians will give different results.  The Poincare recurrence
time is the inverse of the level splitting we have hypothesized
between states on the cosmological horizon.  Thus, if the
Hamiltonian ambiguity is indeed mainly associated with the
description of the horizon states, we expect all of the physics on
the recurrence time scale to be completely unpredictable.
Different Hamiltonians in the universality class will give
different results.  Since, in principle, no actual observations of
this physics can be made, this should not bother us.  Rather, we
might want to view it as a sort of gauge ambiguity in the
description of dS space, which affects mathematical aspects of the
formalism, without affecting the predictions for observable
physics.

It is tantalizing to try to associate this ambiguity with the
gauge invariance of general relativity under change of time
coordinate, the famous Problem of Time.  Indeed, in spaces without
asymptotic boundary, a generally covariant theory does not give
any definite prescription for what the time evolution operator is.
Wheeler-deWitt quantization suggests instead that a system may
have many non-commuting time evolution operators associated with
different semiclassical clocks.  The mutual quantum
incompatibility between different semiclassical clocks is at the
root of the principle of observer Complementarity.   At the
classical level, we have tried to remove this ambiguity for dS
space by choosing the proper time of a given timelike observer to
define the Hamiltonian.  However, a fixed timelike observer is a
classical concept.  Perhaps the inevitable imprecisions we have
discovered in the quantum mechanics of dS space can be related to
a quantum version of the Problem of Time.

There is one final note about measurement theory in dS space,
which connects this discussion to our previous remarks about
symmetry generators.  The freely falling devices we have been
thinking about up to this point do not really correspond to
measurements made by an observer bound to the experiment which
defines the particular static coordinate system that our quantum
formalism refers to.  Rather, they are the best dS approximation
to ``S-matrix meters".  They measure amplitudes which will become
the scattering matrix in the $\Lambda \rightarrow 0$ limit.

Actual measurements done by a static observer are of necessity
less precise than these S-matrix measurements.  Since he remains
bound to the experiment, the size of device that he can build
without gravitationally interacting with the experiment and
changing its result, is much more limited.   In order to read the
results of the S-matrix meters he must send devices out to their
position, which must then accelerate back to him.  These devices
will be affected by the very high temperature radiation that an
accelerated observer experiences near the horizon.

The approximate Poincare generators will have a natural action on
the states measured by the freely falling S-matrix meters.   On
the other hand, the measurements made by the bound observer will
be naturally described in terms of the static dS Hamiltonian.

\section{\bf A toy model of dS quantum mechanics}

This model has been constructed in collaboration with B. Fiol. It
is definitely work in progress, and a lot more progress needs to
be made.  There are several basic principles that we used.  The
first was to realize the spherical geometry of the cosmological
horizon, in a way that was compatible with having a finite number
of states.  This motivates the introduction of fuzzy spheres (M.
Li\cite{li} has utilized fuzzy spheres for a hypothetical
description of dS quantum mechanics.). For the moment our
considerations are restricted to four spacetime dimensions. The
corresponding fuzzy sphere is two dimensional and this is the
only case where a complete technology exists. The restriction to
four dimensions may be only technical, but it may have a deeper
significance.  If the $\Lambda \rightarrow 0$ limit of the theory
is supersymmetric, it must be four dimensional. Only minimal four
dimensional SUGRA admits a dS deformation.

The second principle that we use is the approximation of
Asymptotic Darkness.  That is, we attempt to describe a quantum
theory with stable black holes, and account for the entropy and
energy of these black holes.   The idea is to find a description
of the high energy spectrum, where Hawking decay is negligible.
This should make sense for asymptotically small $\Lambda$.  In dS
space, in contrast to asymptotically infinite spaces, one must,
even in the asymptotic darkness approximation, take into account
the huge reservoir of dS vacuum states.   The asymptotic darkness
approximation also neglects the splittings between black hole
eigenstates, as well as those between vacuum eigenstates.

There is a peculiar feature of the asymptotic darkness
approximation in dS space.  In AdS space, large black holes are
stable.  In asymptotically flat space, they are unstable but
correspond to long lived resonances in scattering amplitudes. The
black hole mass thus has significance even when corrections to
the asymptotic darkness approximation are taken into account. By
contrast, in dS space a black hole decays into objects which fall
through its cosmological horizon\footnote{Even if the black hole
leaves behind a stable remnant, its mass will be much smaller
than that of the hole.  In this case some of the statements below
will be modified, but only by replacing vacuum state by stable
remnant state in appropriate places.}.  Thus, in the full theory,
a black hole must be viewed as a state which can be written as a
superposition of vacuum eigenstates.   It's energy cannot be much
above the dS temperature.   Thus, corrections to the asymptotic
darkness approximation are large.

One can get an intuitive idea for why this might be so by
considering moving black holes in the asymptotic darkness
approximation.  Consider a pair of black holes, which are not
bound by their mutual gravitational attraction, as viewed from
the static frame defined by one of them.  The second black hole
will fall into the cosmological horizon of the first and
therefore has (approximately) zero energy as measured by the
Hamiltonian in the static frame.  This is true even for black
holes which are for some finite range of time, very close to the
central one.   One concludes that there must be superpositions of
vacuum eigenstates whose spacetime description is arbitrarily
close to that of the static frame black hole.  The number of
these states is much larger than the number of static black hole
states\footnote{This discussion is valid for black holes whose
size does not scale to infinity with the dS radius.   The concept
of multiple black holes moving with respect to each other
probably does not make sense close to the Nariai limit.}. It is
easy to imagine that when we split the Hamiltonian as $H = H_{AD}
+V$, in the asymptotic darkness approximation, that the
perturbation $V$ will have order one matrix elements between the
static black hole state and superpositions of vacuum states which
represent close by, moving, black holes.   These can lead to a
significant lowering of the actual black hole eigenvalue.

Given this remark, one may question the utility of the asymptotic
darkness approximation for studying dS space.   Recall however
that there are two interesting Hamiltonians to construct in dS
space with small $\Lambda$.  The other one is the approximate
Poincare Hamiltonian.  The splitting of the Hilbert space into
black hole states and vacuum states will definitely be useful for
the Poincare generator.  Although I will not discuss the Poincare
generator here , this is the best we can do at present, so let us
proceed.

Our fundamental variable will by a complex $N \times N+1$ matrix,
$\Psi_i^A$.  We view it as a bimodule over the fuzzy sphere by
allowing the appropriate irreducible representation of $SU(2)$ to
act on it both on the left and the right.  It is clear that
$\Psi$ transforms in a half integral spin representation. In the
limit $N\rightarrow\infty$ it will be a section of the spinor
bundle on the sphere.  We will quantize $\Psi$ as a fermion,
consistent with the spin statistics theorem,

\eqn{acr}{[\Psi_i^A, (\Psi^{\dagger})^j_B ]_+ = \delta_i^j
\delta^A_B.}

The Fock space formed by these fermionic operators has dimension
$2^{N(N + 1) }$.  Recalling that the radius of the fuzzy sphere
scales like $N$, we see an entropy that scales like the area, at
least for the completely uncertain density matrix on this space.

In the asymptotic darkness approximation we expect the entire
Hilbert space to decompose into eigenspaces of an approximate
Hamiltonian, corresponding to the vacuum, and to black holes of
various masses.  We will take the Hamiltonian in this
approximation to commute with the total fermion number (we do not
expect such a quantum number in the exact theory).  It is then
natural to choose the vacuum density matrix to be the projection
operator on half-filled states, relative to the Fock vacuum of
$\Psi_i^A$.

A corresponding guess for the black hole states is to write $N =
N_+ + N_-$, with $N_+ \geq N_-$.  The black hole density matrix
is then the projection on states, where we only allow filling by
creation operators from either the first $N_+$ rows and $N_+ + 1$
columns, or the last $N_-$ rows and last $N_- +1$ columns to act
on the vacuum, and consider both subsystems to be at half filling.

The microcanonical entropy of this state is ${{N_+^2}\over 2} +
{{N_-^2 }\over 2}$ (for large values of the two integers).   We
identify the two terms in this formula with the entropies of the
cosmological and black hole horizons of the
Schwarzschild-deSitter black holes. They coincide for the maximal
black hole horizon area, which occurs at $N_+ = N_-$. The total
entropy is then equal to one half what we have identified as the
entropy of empty dS space.

Although this is qualitatively the behavior we expect from the
semiclassical thermodynamics of dS space, one would like to do
better and get the relative coefficient in the entropy on the nose
(the absolute coefficient will just be the identification of
Newton's constant in this system).  The cosmological and black
hole horizons satisfy the classical relations

\eqn{horel}{R_+^2 + R_-^2 + R_+ R_- = R^2}

which gives a factor of $2/3$ between the total Nariai black hole
entropy and the empty dS entropy.  Our calculation is clearly
missing an entropy of order $R_+ R_-$.

There are two possible explanations for this discrepancy. First,
we should really be calculating a thermal entropy in the
canonical ensemble at the dS temperature.  We have defined black
hole states by forbidding the excitation of the off diagonal
operators entirely.  Perhaps instead they should be allowed, but
with Boltzmann suppression.  Since the log of the number of these
states is of order $N_+ N_-$ one can hope to make up our entropy
deficit by including them.  In order to get a finite fraction of
the state counting entropy, the energy we assign to these states
has to be of order the dS temperature.

Another possibility is that the factor of $4/3$ between the two
answers should be viewed as an artifact of the asymptotic
darkness approximation, analogous to the factor that occurs in
the free field calculation of the entropy of near extremal black
three branes.  In this view, only the fully interacting theory
will get the coefficient correct.  The fully interacting theory
will however have to deal with the fact that the black holes are
unstable.  In such a calculation, the entropy of empty dS space
should be counted as the thermal entropy of the entire Hilbert
space, including the sectors that are black hole eigenstates in
the asymptotic darkness approximation.  It is not clear to me
whether the semiclassical calculation of Gibbons and Hawking
refers to the full entropy, or the entropy of the empty dS vacuum
in the approximation in which black holes are stable eigenstates,
orthogonal to the vacuum states .

As noted in the beginning of this section, the explicit model of
dS quantum mechanics is as yet in a very primitive stage.
Nonetheless, it gives a hint about the way in which a consistent
quantum theory could reproduce the semiclassical thermodynamics
of dS space.

\section{\bf Conclusions}

Semiclassical analysis leads to the conclusion that a quantum
theory of dS space should have a finite number of states.  This
is implied both by (as yet non-rigorous) arguments that the phase
space of quantum gravity with past and future asymptotically dS
boundary conditions, is compact, and by the combination of the
finiteness of the Gibbons-Hawking entropy and the cutoff on
static energies implied by the existence of a maximal mass black
hole.

The dS entropy is thermal, but analysis of states in dS space
leads to the conclusion that it must primarily represent a very
dense spectrum of levels of the static Hamiltonian at energies
below the dS temperature.  The entropy is then, approximately the
logarithm of the number of these states.  The entropy of states
that can be described by local field theory in a given horizon
volume is bounded by something of order $R^{3/2}$.  This is
consistent with a dual description of the full set of states in
terms of $R^{1/2}$ commuting copies of the field theoretic
degrees of freedom.  I argued that this is approximately the same
as the description of (cutoff )local field theory in global
coordinates, except that the latter formalism implies an infinite
number of copies of the static patch degrees of freedom.   From
the point of view of the static observer, the states
corresponding to local excitations outside his horizon are viewed
as very low energy states on the horizon.  The global field
theoretic picture breaks down drastically when processes which
create horizon scale black holes in a single static patch are
considered.  These put the system into a low entropy state in
which the dynamics outside the horizon is frozen.

The existence of this dense spectrum of levels suggests a
mechanism for understanding the temperature of dS space.  It is
simply the result of interaction of the localizable states with
these low energy horizon degrees of freedom.  The temperature is
an indication of the energy cutoff on the horizon states.
Calculations to verify this conjecture and understand the precise
relation between temperature and cutoff are in progress.

The finiteness of the number of states and the paucity of states
that can be described by field theory inside the cosmological
horizon, puts fundamental limits on measurements in dS space.  In
particular, no self consistent measuring device can be
constructed in the theory, which will retain its classical
character over times comparable to the Poincare recurrence time.

In my view this represents a fundamental ambiguity in the
mathematical description of dS space.  Many mathematical theories
will give the same results for all measurable quantities within
the limits set by the unavoidable lack of precision of
measurement in this system.  The ambiguities are smaller than any
power of the cosmological constant in Planck units, and have
little practical significance, but they are conceptually
important. To someone in a pretentious frame of mind, they
represent the fundamental limit on the basic assumption of
theoretical physics, that the observer can be separated from the
object it observes.  It is likely that most of the ambiguity
refers to the dynamics of the horizon states.  I would conjecture
that they can be described by a more or less random Hamiltonian,
subject to a few constraints.

The above discussion was relevant to the observations made by a
timelike observer in dS space.  The quantum mechanics of such an
observer uses the static dS Hamiltonian.  I showed that in the
limit of vanishing cosmological constant, we should expect the
system to exhibit a new symmetry group, the Poincare
group\footnote{I have argued elsewhere that actually the full
Super-Poincare group will arise in the limit.} (most of) which is
not related to the dS generators, and in particular, not to the
static Hamiltonian. The complicated horizon states completely
decouple from this limiting dynamics.  It describes observations
made by freely falling detectors, near the cosmological horizon.
In the limit, the horizon becomes null infinity and we find the
dynamics of an asymptotically flat space-time.

%%%%%%%%%%%%%%%%%%%%%%%%%%%%%%%%%%%%%%%%%%%%%%%%%%%%%%%%%%%%%%%%%%%%%%%%%=
%%%
%                      REFERENCES                                        =
%
%%%%%%%%%%%%%%%%%%%%%%%%%%%%%%%%%%%%%%%%%%%%%%%%%%%%%%%%%%%%%%%%%%%%%%%%%=
%%%

%\newpage

\end{document}